\documentclass{aip-cp}

\usepackage[numbers]{natbib}
\usepackage{rotating}
\usepackage{graphicx}
\usepackage{mathrsfs}
\usepackage{amssymb}
\usepackage{graphicx}
\usepackage[normalem]{ulem}
\usepackage[dvips]{color}
\usepackage{bm}
\usepackage{longtable}
\usepackage{tabularx}
\usepackage{times}

\renewcommand\sout{\bgroup \color{red} \ULdepth=-.5ex \ULset}

\renewcommand{\rm}[1]{\textrm{#1}}

\begin{document}

\title{Effects of $\phi_0$-meson on the EOS of hyperon star in relativistic mean field model }
\author[aff1]{S. K. Biswal\corref{cor1}}
\corresp[cor1]{Corresponding author and speaker: S. K. Biswal}
\affil[aff1]{CAS Key Laboratory of Theoretical Physics,
 Institute of Theoretical Physics, Chinese Academy of Sciences, Beijing 100190, China}
\maketitle

\begin{abstract}
Nuclear effective interactions are considered as a vital tool
to guide into the region of the high degree of isospin asymmetry and density.
We take varieties of parameter sets of RMF model to show the
parametric dependence of the hyperon star properties.
We add $\phi_0$-meson to $\sigma$-$\omega$-$\rho$ model.
 The effects of $\phi_0$-meson on the equation of state and consequently
on the maximum  mass of the hyperon star are discussed. Due to the inclusion of $\phi_0$-meson the threshold  density of different
hyperon production shift to higher density region. The effects of the
hyperon-meson coupling constants on the maximum mass and radius  of
the hyperon stars are discussed.
\end{abstract}

\section{INTRODUCTION}\label{sec2} 

Nature of the nuclear force under extreme conditions of isospin asymmetry and
 baryon density can be understood from the study of the neutron star properties 
\cite{baym79,latt04,anto13}. Up-gradation of recent experimental techniques can 
only create the nuclear matter up to a few times of nuclear saturation density~\cite{latt04}. 
Due the lack of experimental facilities to probe into the high-density environment, a neutron  star is considered as a solo natural laboratory, which can provide some information about the 
nature of the nuclear force under high density. The global properties of a neutron star carry the information about the nature of equation of state, 
in other words, the nature of the nuclear interaction. Since the last few decades, the limit on 
the maximum mass of the neutron star remains a hot topic for both the nuclear and astrophysicists. 
Theory of general relativity constraints the maximum mass of a neutron star is about 3$M_\odot$~\cite{clif74},
 while the lowest observed neutron star mass is approximately 1.1$M_\odot$ \cite{weis01,mart15}.
 A neutron star is considered as the densest object of the visible universe having 
central density 5-10 times the saturation density~\cite{latt04,baym79}. This high-density 
creates ambiguity about the internal composition of the neutron star. The internal structure 
of a neutron star is not  composed of only nucleons (proton and neutron ) and lepton, as we 
consider in a simple model. From a simple energetic point of view, we can argue that at a high 
density when the Fermi energy of the nucleon crosses the rest mass of the hyperon, there is a 
possibility of conversion of nucleon to hyperon. Usually, the hyperons are produced 
at 2--3 times the saturation density and a neutron star contains 15--20\% 
of the hyperon inside the core \cite{glen85}.  But the 
production of the hyperons reduce the maximum mass of a neutron star \cite{amba60} and 
many calculations can not reproduce the recent observation of neutron star mass about 
2$M_\odot$\cite{anto13,demo10,frei08}. This problem is quoted as hyperon puzzle \cite{amba60}. Primarily, there are three ways to solve this problem :  (a) repulsive hyperon-hyperon interaction through the exchange of vector meson 
\cite{weis12,weis14,oert15} (b) addition of repulsive hyperonic three 
body force ~\cite{vida11,yama13} (c) possibility of phase transition 
to deconfinment quark matter~\cite{ozel10,bona12}.
Still, hyperon puzzle is an open problem, which can be solved by 
knowing hyperon-hyperon interaction in detail. The hyperon-hyperon 
interaction strength plays a major role in deciding the maximum mass and 
other properties of a hyperon star. So it is necessary to have a proper 
investigation for the  effects of hyperon-hyperon interaction 
strength on the various properties. In present contribution, 
I study the effects of the $\phi_0$-meson on the EOS and mass-radius profile 
of the hyperon star with various parameter sets of the relativistic mean field (RMF) model. These 
parameters sets are G1 \cite{frun97}, G2 \cite{frun97},
 IFSU \cite{fatt10}, IFSU* \cite{fatt10}, FSU \cite{todd05},
FSU2 \cite{weic14}, TM1 \cite{suga94}, TM2 \cite{suga94}, PK1 \cite{long04},
 NL3 \cite{lala97}, NL3* \cite{lala09}, NL3-II \cite{lala97},
NL1 \cite{rein86}, NL-RA1 \cite{rash01}, 
  SINPA \cite{mond16}, SINPB \cite{mond16},
GM1 \cite{glen91}, GL97 \cite{glenb97}, GL85 \cite{glen85}, L1 \cite{wale74},
L3 \cite{furn87}, and HS \cite{horo81}. Prespective 
of the taking so many parameter sets is to show the predictive capacity 
of RMF model to reproduce the maximum mass of the hyperon star. 
This proceeding is organised as follows :  in Sec. II, I give a short 
formalism of RMF model and various equations to calculate energy density 
and pressure density, which constitute the equation of state. Tall-Mann Oppenheimer Volkoff equation used to calculate mass and radius of a 
hyperon star. Sec. III, is devoted to discuss the results. In Sec. IV, 
 a summary of the results is given.
\vspace{0.5cm}
\begin{figure}
\centering
\includegraphics[width=9.5cm]{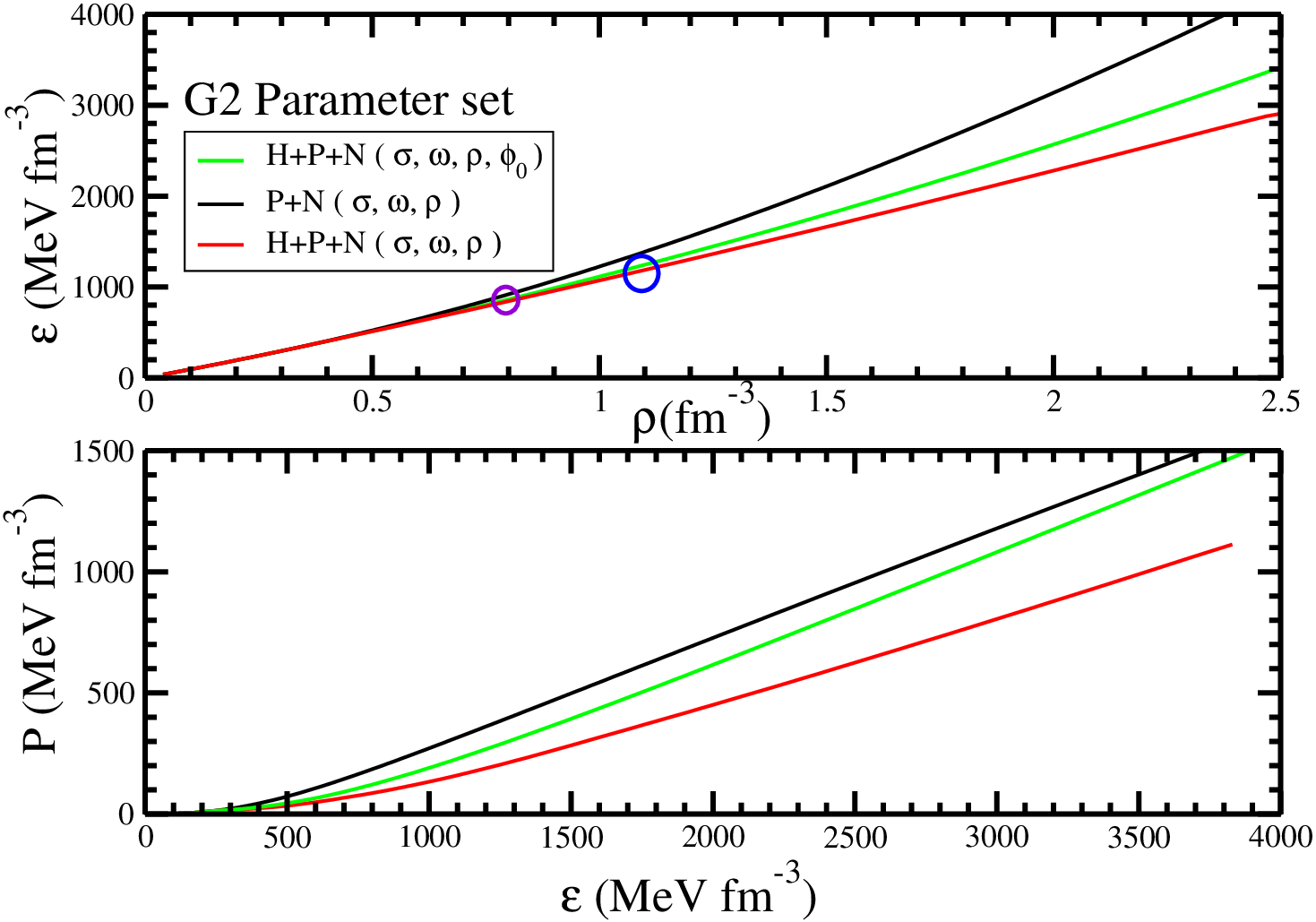}
\caption{The upper panel of the graph shows the variation of the baryon density with energy density. The upper most curve gives the EOS for nucleonic matter (nucleon+lepton), the middle one indicates hyperonic (hyperon+lepton) matter with $\phi_0$-meson contribution. The lower one is same as middle one except contribution of $\phi_0$-meson. The lower panel of the graph shows the variation of the energy density with pressure density. G2 parameter set is used for these calculation.}
\label{fig1}
\end{figure}

\section{Theoretical formalism}\label{form}		
{ Relativistic mean filed model provides a smooth road to go 
from finite 
nuclear system to neutron star system, which has an extreme dense and high 
isospin asymmetry environment. Now-a-days RMF model is used to 
study various properties of the neutron and hyperon star. Starting 
point of the RMF model is an effective Lagrangian. For the present 
calculation, I use an effective Lagrangian which contains non-linear 
interactions of $\sigma$ and $\omega$-meson and cross-coupling of 
various effective mesons~\cite{rein86,mill72,furn87,ring96} },

\begin{eqnarray}
&&{\cal L}=\sum_B\overline{\psi}_B\bigg(
i\gamma^{\mu}\partial_{\mu}-m_B+g_{\sigma B}\sigma -g_{\omega B}\gamma_\mu
 \omega^ \mu  
-\frac{1}{2}g_{\rho B}\gamma_\mu\tau\rho^\mu-g_{\phi_0 B}  \gamma_{\mu} {\phi_0}^{\mu} \bigg)
\psi_B + \frac{1}{2}\partial_{\mu}\sigma\partial_{\mu}\sigma \nonumber \\
&&-m_{\sigma}^2\sigma^2
\left(\frac{1}{2}+\frac{\kappa_3}{3!}\frac{g_{\sigma}\sigma}{m_B}
+\frac{\kappa_4}{4!}\frac{g_{\sigma}^2\sigma^2}{m_B^2}\right)
 - \frac{1}{4}\Omega_{\mu\nu}\Omega^{\mu\nu} 
+\frac{1}{2}m_{\omega}^2
\omega_{\mu}\omega^{\mu}\left(1+\eta_1\frac{g_{\sigma}\sigma}{m_B}
+\frac{\eta_2}{2}\frac{g_{\sigma}^2\sigma^2}{m_B^2}\right)
-\frac{1}{4}R_{\mu\nu}R^{\mu\nu} \nonumber\\
&&+\frac{1}{2}m_{\rho}^2
R_{\mu}R^{\mu}\left(1+\eta_{\rho}
\frac{g_{\sigma}\sigma}{m_B} \right)
+\frac{1}{4!}\zeta_0 \left(g_{\omega}\omega_{\mu}\omega^{\mu}\right)^2 
+\frac{1}{2} {m_\phi}^2 {\phi_\mu}{\phi^\mu}+\sum_l\overline{\psi}_l\left(
i\gamma^{\mu}\partial_{\mu}-m_l\right)\psi_l 
+\Lambda_v R_{\mu}R^{\mu}
(\omega_{\mu}\omega^{\mu}),
\end{eqnarray}
where $\omega_{\mu\nu}$ and $R_{\mu\nu}$ are field tensors
for the $\omega$ and $\rho$ fields respectively and are defined as 
$\omega_{\mu\nu}=\partial_\mu \omega_\nu-\partial_\nu \omega_\mu$ and
$R_{\mu\nu}=\partial_\mu R_\nu-\partial_\nu R_\mu$. The symbols are
carrying their usual meanings.
{$\sigma$, $\omega$ and $\rho$-meson are exchanged between the nucleons, 
 while $\phi_0$ being a strange meson it is exchanged between the 
 hyperons only. The coupling constants of nucleon-meson
interactions are fitted  to reproduce the desired
nuclear matter saturation properties and finite nuclear properties, like 
charge radius, binding energy, and monopole excitation energy of
a set of spherical nuclei.} The nature of the interaction depends on the
quantum numbers and masses of the intermediate mesons. $\sigma$ (T=0, S=0) is an isoscalar-scalar meson, it gives intermediate attractive interaction.
$\omega$-meson (T=0, S=1) is an isoscalar-vector meson, which gives
short-range repulsive interaction. $\rho$-meson (T=1, S=1) is an isovector
vector meson, whose interaction is account for the isospin asymmetry. Newly
added $\phi_0$-meson is a vector meson, which gives similar interaction
like $\omega$-meson \cite{cava08,bipa89,bedn03,spal99,bunt04,scha96,feng92}.
By using classical Euler-Lagrangian equation of motion, we get the various
equation of motions for the different mesons.

\begin{flushleft}
\begin{eqnarray}
m_{\sigma}^2 \left(\sigma_0+\frac{g_{\sigma N}\kappa_3\sigma_0}{2m_B}
+\frac{\kappa_4 g_{\sigma N}^2\sigma_0^2}{6m_B^2} \right) \sigma_0
-\frac{1}{2}m_{\rho}^2\eta_{\rho}\frac{g_{\sigma N}\rho_{03}^2}{m_B}
-\frac{1}{2}m_{\omega}^2\left(\eta_1\frac{g_{\sigma N}}{m_B}
+\eta_2\frac{g_{\sigma N}^2\sigma_0}{m_B^2}\right)\omega_0^2
=\sum g_{\sigma B}\rho^s_B.
\end{eqnarray}
\end{flushleft}
\begin{flushleft}
\begin{eqnarray}
m_{\omega}^2\left(1+\eta_1\frac{g_{\sigma N} \sigma_0}{m_B}
+\frac{\eta_2}{2}\frac{g_{\sigma N}^2\sigma_0^2}{m_B^2}\right)\omega_0 
+\frac{1}{6}\zeta_0g_{\omega N}^2\omega_0^3
=\sum g_{\omega B}\rho_B.
\end{eqnarray}
\end{flushleft}
\begin{eqnarray}
 m_{\rho}^2\left(1+\eta_{\rho}\frac{g_{\sigma N}\sigma_0}{m_B}\right)R_0
=\frac{1}{2}\sum g_{\rho B}\rho_{B3}.
\end{eqnarray}
\begin{eqnarray}
{m_\phi}^2 \phi_0 =\sum g_{\phi B} \rho_B.
\end{eqnarray}

Equation for the $\phi_0$-meson is similar to the $\omega $-meson
except the coupling constants. 
We use the  expression,
\begin{eqnarray}
U_{Y} = {m_n}(\frac{m_n^*}{m_n}-1)x_{\sigma Y}+(\frac{g_\omega}
{m_\omega})^2\rho_0 x_{\omega Y},
\end{eqnarray}
 to calculate the hyperon potential depth.
 Y stands for the different hyperons ( $\Lambda, \Sigma, \Xi $ ).
$x_{\sigma Y}$ and $x_{\omega Y}$ are the coupling constants of the
hyperon-meson interactions and $\rho_0$ is the saturation density.
We  choose ${U_{\Lambda}}^{(N)}=-30$ MeV
\cite{batt97,mill88},
${U_{\Sigma}}^{(N)}= +40$ MeV \cite{frie07},
 and ${U_{\Xi}}^{(N)}= -28 $ MeV \cite{glen91}. The hyperon-meson
coupling constants $x_{\sigma Y}$ and $x_{\omega Y}$  are fitted in a such
 a way that hyperon potential depth for various hyperons can be reproduced.
We can vary the
$x_{\sigma Y}$ and $x_{\omega Y}$ for different combinations to get the
 depth of the hyperon potentials. The  hyperon interaction strengths with $\rho$-mesons are fitted
according to the SU(6) symmetry \cite{dove84}, $x_{\Lambda\rho}=$ 0, $x_{\Sigma\rho}=$ 2, $x_{\Xi\rho}=$1.
The interaction strengths of the hyperons with $\phi_0$-meson  are given by
$x_{\phi\Lambda}=$$-{\sqrt{2}}/{3}$ g$_{\omega N}$, $ x_{\phi\Sigma}=
-{\sqrt{2}}/{3}$ $g_{\omega N}$, $x_{\phi\Xi}=-{2\sqrt{2}}/{3}$ $g_{\omega N}$.
These equations form a set of self-
consistent equations, which can be solved by iterative method to find
various meson fields and densities. Using energy-momentum tensor, the total
energy and pressure density can be written,
\begin{eqnarray}\label{energy}
&&{\cal E}=\sum_B\frac{Y_B}{(2\pi)^3}\int_0^{k_F^B} d^3k \sqrt{k ^2+{m_B}^*}
 + m_{\sigma}^2\sigma_0^2\left(\frac{1}{2}+\frac{\kappa_3}{3!}
\frac{g_{\sigma}\sigma_0}{m_B}+\frac{\kappa_4}{4!}
\frac{g_{\sigma}^2\sigma_0^2}{m_B^2}\right) 
 + \frac{1}{2}m_{\omega}^2 \omega_0^2\left(1+\eta_1
\frac{g_{\sigma}\sigma_0}{m_B}+\frac{\eta_2}{2}
\frac{g_{\sigma}^2\sigma_0^2}{m_B^2}\right) \nonumber \\
&& + \frac{1}{2}m_{\rho}^2 \rho_{03}^2\left(1+\eta_{\rho}
\frac{g_{\sigma}\sigma_0}{m_B} \right)+\frac{1}{2}{m_\phi}^2 \phi^2
+\frac{1}{8}\zeta_0g_{\omega}^2\omega_0^4
+\sum_l \int_0^{k_F^l} \sqrt{k^2+{m_l}^2} k^2 dk
+3\Lambda_V \omega_0^2 R_0^2,
\end{eqnarray}
and
\begin{eqnarray}\label{pressure}
&&{\cal P}=\sum_B\frac{Y_B}{3(2\pi)^3}\int_0^{k_F^B}\frac{k^2 d^3k}{\sqrt{k^2+{m_B^*}^2}}
- m_{\sigma}^2\sigma_0^2\left(\frac{1}{2}+\frac{\kappa_3}{3!}
\frac{g_{\sigma}\sigma_0}{m_B}+\frac{\kappa_4}{4!}
\frac{g_{\sigma}^2\sigma_0^2}{m_B^2}\right) 
 + \frac{1}{2}m_{\omega}^2 \omega_0^2\left(1+\eta_1
\frac{g_{\sigma}\sigma_0}{m_B}+\frac{\eta_2}{2}
\frac{g_{\sigma}^2\sigma_0^2}{m_B^2}\right) \nonumber \\
&& + \frac{1}{2}m_{\rho}^2 \rho_{03}^2\left(1+\eta_{\rho}
\frac{g_{\sigma}\sigma_0}{m_B} \right)+\Lambda_V R_0^2 \omega_0^2 
+\frac{1}{24}\zeta_0g_{\omega}^2\omega_0^4 
+\frac{1}{3\pi^2}\sum_l \int_0^{k_F^l} \frac{k^4 dk}{\sqrt{k^2+m_l^2}},
\end{eqnarray}
where, $l$ stands for the leptons like electron and muon.
 Variation of total energy and pressure density
 with baryon density known as the equation of state of the nuclear matter.
Putting $\beta$-equilibrium  and charge neutrality conditions, it can be
converted to the star matter equation of state. These equation of states
are the inputs of the Tolman-Oppenheimer-Volkoff (TOV) \cite{tolm39,oppe39} equation, which is
given by
\begin{eqnarray}
\frac{\partial P}{\partial r}= - \frac{(P+\rho)(M(r)+4\pi r^3 P)}{r(r-2m)},
\end{eqnarray}
\begin{eqnarray}
\frac{\partial m}{\partial r}=4\pi r^2 \rho(r),
\end{eqnarray}
where $m(r)$ is the enclosed gravitational mass, $P$ is the pressure, $\cal E$
is the total energy density and $r$ is the radial variable.
These two coupled hydro-static equations are solved to get the mass and radius
of the neutron star at a certain central density. Different central density
gives different combination of mass and radius and one particular choice of
central density gives maximum  mass of the neutron star for a given EOS.



\vspace{0.6cm}
\begin{figure}[h]
\vspace{0.6cm}
\centering
\includegraphics[width=9.5cm]{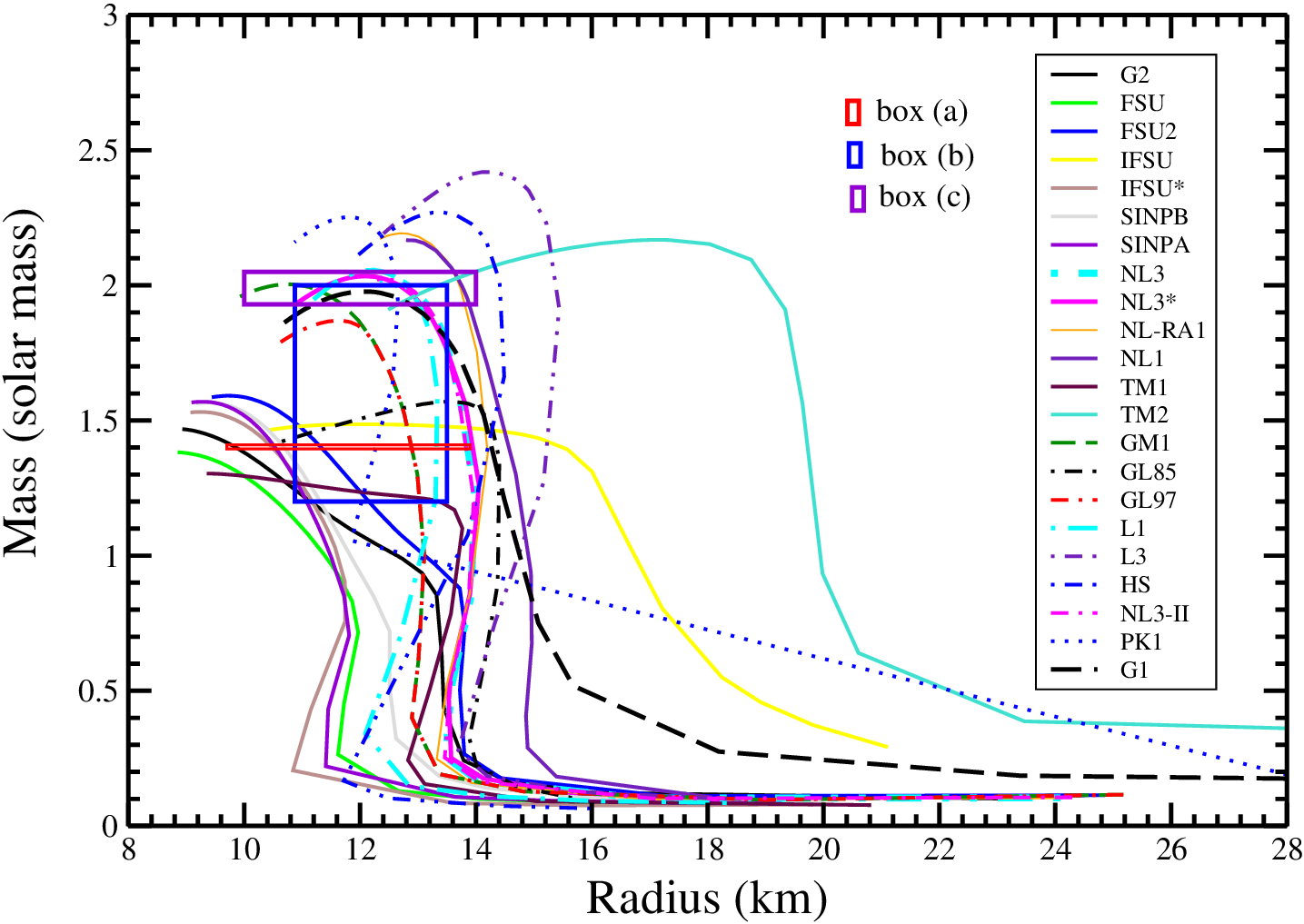}
\caption{ The figure shows the mass-radius curve for the various parameter 
	sets of the RMF model. Box (a), shows acceptable range of the radius 
	of canonical star (1.4$M_\odot$), while box (b) represents the star 
	mass in the range 1.2--2 $M_\odot$ with radius 10.7--13.5 km. The box 
	(c) shows the limit on the maximum mass of the neutron star ie. 1.93--
	2.05 $M_\odot$.
	}\label{fig2}
\end{figure}

\begin{figure}[h]
\vspace{0.6cm}
\centering
\includegraphics[width=9.5cm]{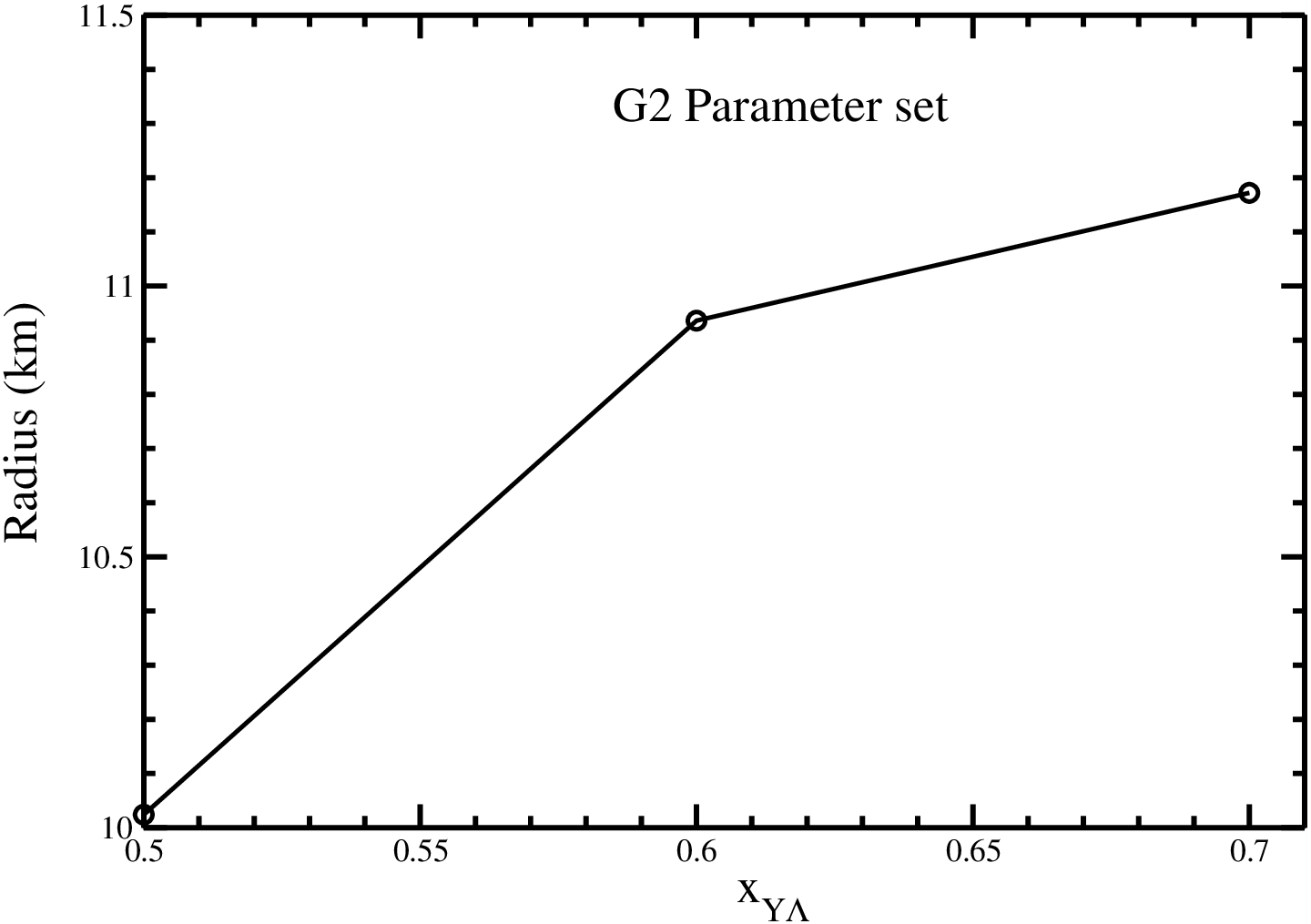}
\caption{ The graph shows the variation of the radius of the canonical star 
	(1.4$M_\odot$) with hyperon-meson coupling constants $x_{\Lambda Y}$.
        }\label{fig3}
\end{figure}

\section{Results and discussions}\label{result}

In the present proceeding, I use 22 parameter sets of the RMF model 
to calculate the properties of the hyperon star. These parameter sets
 are divided into five groups according to the nature of the nucleon-nucleon 
 interaction. The parameter sets belong to a same group
 are only different from each other by the value of the coupling constants 
and saturation properties. From example, group I, contain G1 and G2 parameter
 sets. Both G1 and G2 have a similar type of nucleon-meson interaction 
and contain the same cross-couplings and self-interactions among mesons. These 
22 parameter sets are commonly used in RMF calculations. These parameter sets 
are differentiated from each other in a wide range of saturation properties 
like incompressibility (K), symmetry energy (J), saturation binding energy 
(E/A) and saturation density ($\rho_0$). These saturation properties also cover a wide range of value such as, incompressibility of NL1 is 211.4 MeV, while that of L1  
is 626.3 MeV. Similarly, symmetry energy value ranges from 21.7 MeV (L1) to
 43.7 MeV (NL1).  These two quantities of nuclear matter affect the 
 EOS of the neutron star in a significant way. The perspective of 
taking so many parameter sets is that to check the predictive capacity of RMF 
 model with different parameter sets. 

Nuclear matter equation of state is considered as one of the most important 
ingredient for the calculation of neutron star properties. Before the 
discussion the effects of the $\phi_0$-meson on the properties 
of the hyperon star, 
it is wise to investigate how the $\phi_0$-meson affects the EOS. 
In Fig.~\ref{fig1} the effects of $\phi_0$-meson on the EOS of
neutron star matter
are shown. The upper panel shows the variation of the baryon density with
energy density. Upper curve of the upper panel is for the pure neutron-proton matter with no contribution of the hyperons. This curve is the stiffest one.
The lowest curve contains the contribution of hyperon. The middle one contain
contribution of the hyperon along with the $\phi_0$-meson as 
inter-mediating meson.
The graph clearly shows the $\phi_0$-meson makes the EOS stiff. So it increases
the maximum mass of the hyperon star. In the upper panel
of the Fig.~\ref{fig1},
contribution of the $\phi_0$-meson comes around 0.8 fm$^{-3}$, which is shown by 
a blue circle. So it makes
the EOS with $\phi_0$ and without $\phi_0$-meson deviate from each other around 0.8 fm$^{-3}$.
The lower panel shows the variation of the energy density with pressure density.
The pressure-energy density graph also follows similar trend as in the 
upper panel. The hyperon star matter without $\phi_0$-meson shows a soft EOS, 
while with $\phi_0$-meson it shows a comparative stiff EOS.
In Fig.\ref{fig2}, I show the mass-radius graph with different parameter sets. 
Three boxes are shown in the figure. The box (a) represents the radius of the 
canonical star (1.4 $M_\odot$) \cite{hebe10}. From the study of chiral 
effective model, authors in Ref.\cite{hebe10} suggested that the radius 
of the canonical star lies 
in the range 9.7--13.9 km. The figure shows that most of the parameter sets 
are unable to reproduce the radius of the canonical star (1.4 $M_\odot$) 
in the above range. Only a few parameter sets lik FSU2, PK1, GL85, Gl97
, SINPA, SINPB, NL3, NL3*, IFSU*, and G2  can give the radius of the canonical star in the above range. 
But the radius of the canonical star depends on the hyperon-meson coupling 
constants. For example, if I change the hyperon-meson coupling constants 
, while keeping fix the different hyperon potential, the radius of the 
canonical star increases monotonically with hyperon-meson interaction strength.
The box (b), shows the star with mass 1.2--2 $M_\odot$ and radius in the 
range 10.7--13.5 km. Many parameter sets are able to reproduce the mass 
and radius in this range. These parameter sets are GL97, GL85, FSU, TM1, 
PK1, GM1 SINPA, and SINPB. Simillarly, the box (c) indicates the limit of the 
maximum mass of the neutron star from recent observations. GM1, NL3-II, 
NL3, NL3*, parameter sets have maximum mass in this recent 
observation limit, which is 1.93--2.05 $M_\odot$. 
Fig.\ref{fig3}, shows the variation of radius of the canonical star (1.4 $M_\odot$) with hyperon-meson coupling constants. 
For the quantitative check, $x_{\sigma\Lambda}$ changed from 0.5 to 0.7 as 
result the radius changed from 10.0278 km to 11.275 km. The radius of the 
canonical star changes to 13\% by changing the  hyperon-meson coupling constant
to 0.2. While changing the $x_{\sigma\Lambda}$ values, we also keep 
changing the value of 
$x_{\omega\Lambda}$ to fix the value of the $U_\Lambda$ at -30 MeV. 
In the similar way, I take care the hyperon-meson coupling constants 
for other hyperons. I keep the potential of $U_\Lambda$, and 
$U_\Sigma$ and $U_\Xi$, while changing the $x_{\sigma Y}$ and $x_{\omega Y}$.

\vspace{0.8cm}
\begin{figure}[h]
\vspace{0.7cm}
\centering
\includegraphics[width=9.5cm]{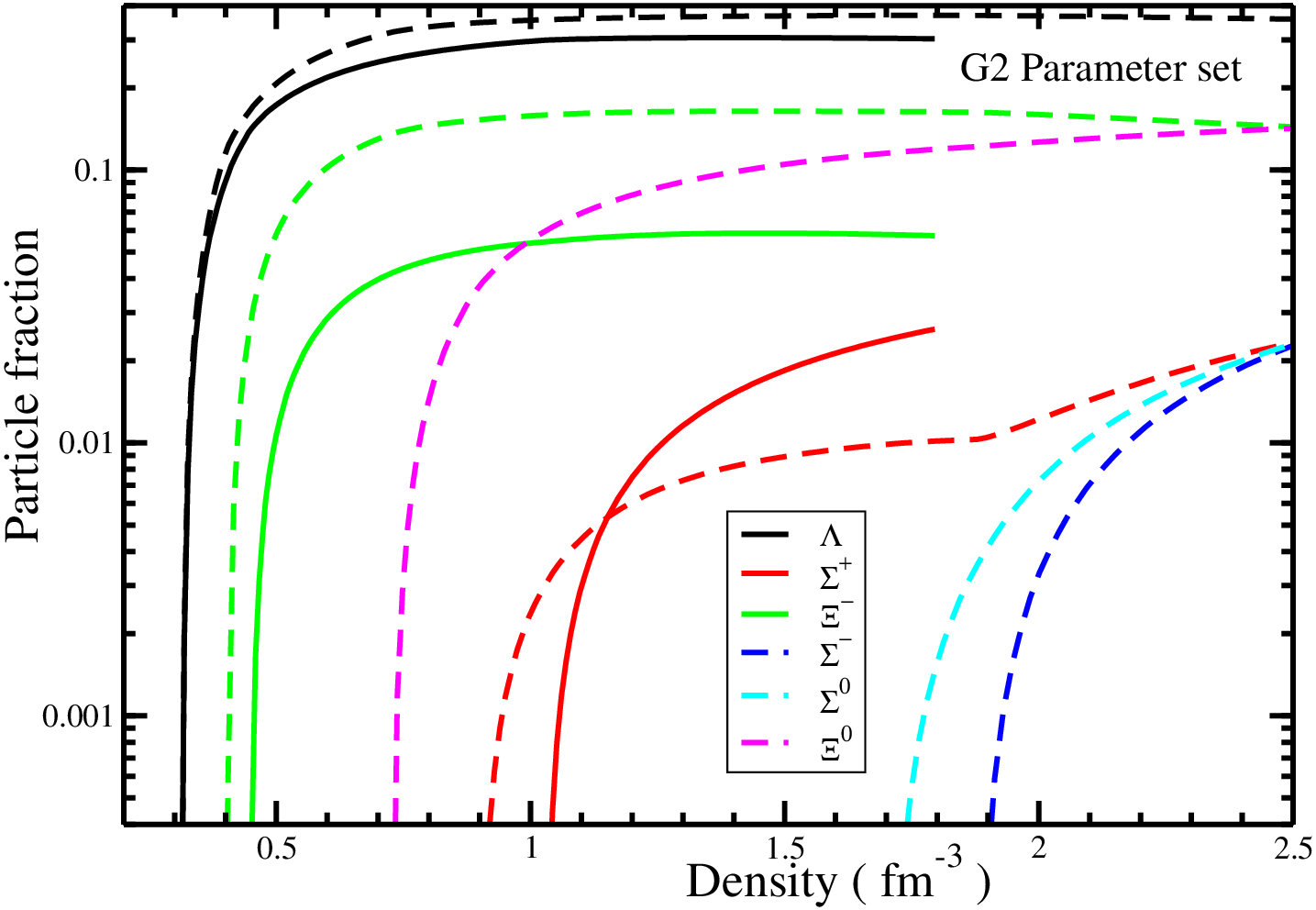}
\caption{Particle fraction of the hyperon with G2 parameter set. The solid lines
show  the particle fraction of different types of hyperons with $\phi_0$-meson and the corresponding dotted lines show the result of hyperons without
$\phi_0$-meson.}
\label{fig4}
\end{figure}

Fig.\ref{fig4}, shows the effect of the $\phi_0$-meson on the hyperon 
production with G2 parameter set. The solid lines represent the hyperon 
production with $\phi_0$-meson 
contribution, while the dotted line without $\phi_0$-meson. This graph 
shows that $\phi_0$-meson shifts the threshold density (density at which 
different hyperons are produced) to higher density. For example, without 
the contribution of the $\phi_0$-meson the $\Xi$ produces at a density 
0.3989 fm $^{-3}$, but the addition of  $\phi_0$-meson  shifts 
the threshold density to 0.461 fm$^{-3}$. Similarly, $\Sigma^+$ hyperon's 
threshold  density  shifts from 0.914 fm$^{-3}$ to 1.038 fm$^{-3}$. 
The threshold  density of the $\Lambda$-hyperon does not shift by 
a significant amount. In the table 1, the maximum mass, the corresponding 
radius, compactness, 
and the central density at which maximum mass occurs for the hyperon 
star are given with different parameter sets. Results are given for the 
neutron star, hyperon star with $\sigma$-$\omega$-$\rho$-model and 
hyperon star with $\sigma$-$\omega$-$\rho$-$\phi_0$-model. We conclude 
from the table 1, for all the parameter sets the maximum masses follow 
a general trend i.e. $M_{\mathrm{max}}^N$ ($\sigma$-$\omega$-$\rho$) $>$
$M_{\mathrm{max}}^H$ ($\sigma$-$\omega$-$\rho$-$\phi_0$) $>$ $M_{\mathrm{max}}^H$ ($\sigma$-$\omega$-$\rho$), where $M_{\mathrm{max}}^N$ ($\sigma$-$\omega$-$\rho$) is the 
maximum mass of the proton-neutron star in $\sigma$-$\omega$-$\rho$ 
model and $M_{\mathrm{max}}^H$ ($\sigma$-$\omega$-$\rho$-$\phi_0$) is the
maximum mass of the hyperon star in $\sigma$-$\omega$-$\rho$-$\phi_0$ model.
But the radius does not follows any common trend in all parameter sets. 
The compactness (M/R) also follows a general trend i.e. $(M/R)^N$ ($\sigma$-$\omega$-$\rho$) $>$ $(M/R)^H$ ($\sigma$-$\omega$-$\rho$-$\phi_0$) $>$ $(M/R)^H$ ($\sigma$-$\omega$-$\rho$) for all the parameter sets. The data shows by adding 
the $\phi_0$-meson the compactness of the hyperon star increases, this is 
mainly due to the increse of the maximum mass with the inclusion of $\phi_0$-meson. The central 
density (${\cal E}_c$) at which the maximum mass occurs, also follows a 
common pattern for all the parameter sets i.e.  ${\cal E}_c^N$ ($\sigma$-$\omega$-$\rho$) $<$ ${\cal E}_c^H$ ($\sigma$-$\omega$-$\rho$-$\phi_0$) $<$ ${\cal E}_c^H$ ($\sigma$-$\omega$-$\rho$).

\vspace{1em}
\begin{table*}[h]
\hspace{3.0 cm}
\centering
	\caption{Maximum mass (M), corresponding radius at maximum mass (R), compactness (M/R), and central density (${\cal E}_c$) for the neutron and hyperon stars are given with various parameter sets. } 
\renewcommand{\tabcolsep}{0.08 cm}
\renewcommand{\arraystretch}{1.}
{\begin{tabular}{|c| c| c| c| c| c| c| c| c| c| c| c| c| c| c|}
\hline
&\multicolumn{4}{|c|}{Neutron star}&\multicolumn{4}{|c|}{Hyperon star
with $\phi_0$}
&\multicolumn{4}{|c|}{Hyperon star without $\phi_0$}\\
\hline
	parameter &M &R  & M/R & ${\cal E}_c\times 10^{15}$  
	& M &R & M/R &${\cal E}_c\times 10^{15}$  &M  &R  & M/R
	&${\cal E}_c\times 10^{15}$ \\
	sets & ($M_\odot$) &(km)& ($M_\odot$/km) & ( g cm$^{-3}$)  
	& ($M_\odot$)&(km)& ($M_\odot$/km) & ( g cm$^{-3}$) 
	& ($M_\odot$)&(km)& ($M_\odot$/km) & (g cm$^{-3}$) \\
\hline
\multicolumn{13}{|c|}{GROUP I}\\
\hline
G2 & 1.938 & 11.126  &0.174&2.317 & 1.576 &9.622&0.163&3.387&1.299&9.054&0.135&3.921 \\
\hline
G1& 2.162 &  12.244 &0.176 & 1.871&1.881&11.712 &0.160&2.139&1.816&12.048&0.150&1.960  \\
\hline
 \multicolumn{13}{|c|}{GROUP II}\\
\hline
FSU& 1.722  & 10.654 & 0.161& 2.495 & 1.419 & 9.280&0.152& 3.743&1.100&8.788& 0.125& 3.921  \\
\hline
FSU2& 2.072 & 12.036  & 0.172& 1.960 & 1.779 & 11.399&0.155&2.139&1.377&10.812&0.127&2.674 \\
\hline
IFSU& 1.898 & 12.612 & 0.150& 1.960& 1.793 & 13.450&0.133&1.604&1.750&13.890 & 0.125& 1.426\\
\hline
IFSU*& 1.985 & 11.386 & 0.174  &1.529  & 1.763&10.90&0.161&2.317&1.570&10.820&
0.145&2.317 \\
\hline
SINPA&2.001 &11.350 &0.176&2.139 &1.750 & 10.652 &0.164&2.495 &1.525& 10.334 &0.147&2.674    \\
\hline
 SINPB& 1.994 &11.468  &0.173&2.139 &1.719&10.536&0.163& 2.674 & 1.404& 10.258&0.136&2.674  \\
\hline
 \multicolumn{13}{|c|}{GROUP III}\\
\hline
TM1&2.176& 12.236  & 0.177&1.871& 1.966&11.986  & 0.164& 1.960& 1.798& 12.228
& 0.1470& 1.782     \\
\hline
TM2& 2.622 & 16.508 & 0.158&  1.069 &2.512   & 17.062 &0.147& 0.944& 2.479&17.358& 0.142& 0.929   \\
\hline
PK1&2.489 & 14.042 & 0.177 &1.604  &2.275& 13.688& 0.166&1.782& 2.128& 13.904&
0.153&1.782    \\
\hline
 \multicolumn{13}{|c|}{GROUP IV}\\
\hline
NL3& 2.774  & 13.154  & 0.210&1.604&2.633 &  13.012& 0.193& 1.604& 2.529& 13.012&0.194& 1.6044      \\
\hline
NL3*& 2.760 & 13.102& 0.210&1.604 & 2.605 & 12.938&0.201 & 1.604&2.500& 12.930& 0.193& 1.604    \\
\hline
NL1& 2.844&13.630 & 0.208&1.426 & 2.653 &13.740&0.193&1.604&2.506&13.190&0.189&1.604  \\
\hline

GM1& 2.370&12.012& 0.197& 1.960& 2.280 & 12.14 &0.187&1.871&2.230&12.21&0.182&1.77   \\
\hline
GL85& 2.168& 12.092&0.220&1.960  &2.122& 12.242&0.173& 1.871&2.106&12.223&0.172&1.871\\
\hline
GL97&2.003 &10.790&0.185&2.495&1.919 &10.832&2.495&0.177&1.881&10.894&0.172&2.495    \\
\hline
NL3-II& 2.774 &13.146 &0.211&1.604  &2.594   & 12.94 &0.200&1.604&2.474&12.742&0.194&1.782  \\
\hline
NL-RA1& 2.783 &13.420 &0.207&1.426 &2.631&13.050&0.201 & 1.604&2.516&13.030&
0.193&1.604    \\
\hline
 \multicolumn{13}{|c|}{GROUP V}\\
\hline
HS& 2.974 &14.176&0.209&1.2478  &2.853&13.848 & 0.206&1.4261&2.770&13.860&0.199&1.426    \\
\hline
L1& 2.744 & 13.004 &0.211& 1.604&2.056  & 12.176 &0.168 &1.871&2.00&12.372&0.161 &1.786  \\
\hline
L3& 3.186 & 15.224 &0.209 &1.069&2.088&12.374&0.168&1.782&1.692&12.294&0.137
& 1.069    \\
\hline
\end{tabular}
\label{tab4}}
\end{table*}

\section{Summary and Conclusions}\label{conc}
In summary, I study the properties of the hyperon stars with the various
parameter sets of RMF model. The predictive capacity of the various
parameter sets to reproduce the canonical mass-radius relationship 
are  discussed with  hyperonic degrees of freedom. 
Out of 22 parameter sets only few parameter sets like FSU2, PK1, GL85, Gl97
, SINPA, SINPB, NL3, NL3*, IFSU*, and G2 can able to reproduce the radius of the canonical star (1.4 $M_\odot$) in the range 9.7 km to 13.9 km. But radius of 1.4 $M_\odot$ star depends on the 
strongly on the hyperon-meson couplings. The radius of a canonical  
star increases monotonically with hyperon-meson interaction in-spite of a fixed 
hyperon potential depth. The radius of a canonical star change by 13\% with 
a small change of 0.2 of the hyperon-meson coupling constants. This shows not only the depth of the hyperon potential but also the range of the hyperon-meson 
coupling constants are improtant. More hyper-nuclei data required to fix the range of the hyperon-hyperon interaction. As the $\phi_0$ is a 
vector meson, so it  gives  
repulsive interaction among the hyperons and makes the EOS comparatively stiff 
The stiff EOS  increases the maximum mass of the hyperon star. The 
compactness of the hyperon star increases with inclusion of the 
$\phi_0$-meson. The $\phi_0$-meson push the threshold density of 
hyperon production to higher density.

\section{Acknowledgement}
All the calculations are preformed in Institute of Theoretical Physics,
Chinese Academy of Sciences. This work has been supported by the National 
Key R\&D Program of China (2018YFA0404402), the NSF of China 
(11525524, 11621131001, 11647601, 11747601, and 11711540016), 
the CAS Key research Program (QYZDB-SSWSYS013 and XDPB09), and 
the IAEA CRP "F41033".


\end{document}